\documentclass[12pt]{article}
\usepackage{cite}
\usepackage{amsfonts,latexsym}
\usepackage{epsfig}
\begin{document}
\newcommand{\de}{\delta}\newcommand{\ga}{\gamma}
\newcommand{\e}{\epsilon} \newcommand{\ot}{\otimes}
\newcommand{\be}{\begin{equation}} \newcommand{\ee}{\end{equation}}
\newcommand{\ba}{\begin{array}} \newcommand{\ea}{\end{array}}
\newcommand{\beq}{\begin{equation}}\newcommand{\eeq}{\end{equation}}
\newcommand{\tmod}{{\cal T}}\newcommand{\amod}{{\cal A}}
\newcommand{\bemod}{{\cal B}}\newcommand{\cmod}{{\cal C}}
\newcommand{\dmod}{{\cal D}}\newcommand{\hmod}{{\cal H}}
\newcommand{\s}{\scriptstyle}\newcommand{\tr}{{\rm tr}}
\newcommand{\einsop}{{\bf 1}}
\def\j{\mathcal{J}}
\def\R{\overline{R}} \def\doa{\downarrow}
\def\dag{\dagger}
\def\ve{\epsilon}
\def\si{\sigma}
\def\ga{\gamma}
\def\no{\nonumber}
\def\le{\langle}
\def\re{\rangle}
\def\lt{\left}
\def\rt{\right}
\def\dwn{\downarrow}
\def\up{\uparrow}
\def\dag{\dagger}
\def\nonum{\nonumber}
\newcommand{\reff}[1]{eq.~(\ref{#1})}

\title{Quantum dynamics of a model for two Josephson-coupled
Bose--Einstein condensates}

\author{A. P. Tonel$^{1}$\footnote{sponsored by CNPq-Brazil},
~J. Links$^{1}$
and A. Foerster$^{2}$
\vspace{1.0cm}\\
$^{1}$ Centre for Mathematical Physics, School of Physical Sciences, \\
The University of Queensland, Brisbane, 4072, Australia
\vspace{0.5cm}\\
$^{2}$ Instituto de F\'{\i}sica da UFRGS, \\
Av. Bento Gon\c{c}alves 9500, Porto Alegre, RS - Brazil}
%\vspace{0.5cm}\\
%\vspace{0.5cm}\\$^{3}$Departamento de F\'{\i}sica da UNESP \\
%Av. Eng. Luiz Edmundo Carrijo Coube, s/n,  Bauru, SP-Brazil
%}

\maketitle

\begin{abstract}
In this work we investigate the quantum dynamics of a model for
two single-mode Bose--Einstein condensates which are coupled via
Josephson tunneling. Using direct numerical diagonalisation of the
Hamiltonian, we compute the time evolution of the expectation value for 
the relative particle
number across a wide range of couplings. Our analysis shows that
the system exhibits rich and complex behaviours varying between 
harmonic and non-harmonic oscillations, particularly around the
threshold coupling between the delocalised and self-trapping phases. 
We show
that these behaviours are dependent on both the initial state of the
system as well as regime of the coupling. In addition, a study of the
dynamics for the variance of the relative particle number expectation
 and the entanglement for
different initial states is presented in detail.

% do not depend on the number of particles
\end{abstract}


\vfil\eject
%%%%%%%%%%%%%%%%%
\section{Introduction}

The phenomenon of Bose--Einstein condensation, while predicted
long ago \cite{bose,eins}, is nowadays responsible for many of the
current perspectives on the potential applications of quantum
systems. This point of view has arisen with the experimental
observation of condensation in systems of ultracold dilute alkali
gases, realised by several research groups using magnetic traps
with various kinds of confining geometries. Reviews of the
breakthroughs that have lead to the current state of development
can be found in \cite{cw,ak}. These types of experimental
apparatus open up the possibility for studying dynamical regimes
at the frontier between the quantum and classical scenarios, where
new macroscopic quantum phenomenon can emerge; for example,
coherent atomic lasers \cite{int3}, and the new chemistry of
atomic-molecular condensates \cite{dctw}. Subsequently,
Bose--Einstein condensates are seen as one of the main tools to
investigate, verify and improve our understanding of many concepts
and principles in quantum physics, such as entanglement \cite{y}.
It is widely accepted that entanglement is the main resource
needed for the implementation of quantum computation \cite{nc}.

The work on ultracold atomic gases 
has demonstrated the occurence of Rabi oscillations
due to quantum tunneling between two internal states of a
condensate, similar to the Josephson effect in superconductors.
Moreover, quite unexpected results were found in
\cite{intaa,intbb}, where it was shown that the temporal evolution
of the expectation value of the number of particles exhibits a
collapse and revival behaviour (see also e.g. \cite{gmehb}). 
This is a novel effect, not observed
with spatial and temporal resolution in other systems where
quantum effects take a macroscopic scale, like the superfluid and
the superconducting systems.

{}From the theoretical point of view tunneling in Bose--Einstein
condensates has been widely investigated using a simple two-mode
Hamiltonian (see eq. (\ref{ham}) in the next section). This model
has been studied by many authors using a variety of methods, such as
the Gross-Pitaevskii approximation \cite{leg}, mean-field theory
\cite{milb,hmm},  quantum phase model \cite{int8}, and the Bethe
ansatz method \cite{lz,zlmg}. Our aim in this work is to expand on
the theoretic knowledge of tunneling in Bose--Einstein condensates
by undertaking a detailed and systematic analysis of the quantum
dynamics of this two-mode Hamiltonian by means of the method of
direct numerical diagonalisation. In particular, we will consider the
temporal evolution of the expectation value for the relative
number of particles between the two condensates 
for different choices of
the coupling parameters and distinct initial states. In this
procedure we will keep the total number of particles fixed, while
the coupling of the Hamiltonian is varied. Using the numerical  
diagonalisation method, we are able to study the model across all
coupling regimes. 

We also implement this same approach to investigate the temporal
evolution of entanglement for this model, and contrast this against the
variance of the relative number of particles. As discussed in
\cite{mjcz}, there are generally two approaches for creating
entanglement in many-body systems. One is to engineer
a gapped system whose ground
state is known to be entangled. By then sufficiently cooling the system to 
the ground state configuration, it will be entangled.
In this respect there have been studies of ground state entanglement for
the $XYh$ model \cite{on,oaff}, the BCS model \cite{md} and two-mode
Bose-Einstein condensates including the present model \cite{hmm}.
Alternatively, one can manipulate a system so that 
a given initial state temporally evolves into an entangled state.  
From this point of view entanglement dynamics for coupled Bose-Einstein
condensates have been investigated in \cite{saf}, using the same model
we study here, but with severe constraints on the choice of couplings,
and also in \cite{hmm}. Here we will investigate the behaviour of the 
entanglement evolution across different coupling regimes.  

%The paper is organised in the following way. In section II we
%present the model and the procedure for investigating the dynamics
%of the Hamiltonian. In section III we discuss the expectation
%value for the relative number of particles 
%for different initial states. In section IV we compute the
%time evolution for the variance of the relative particle number
%and the entanglement, using three different initial states.
%Section V is reserved for our conclusions.

%%%%%%%%%%%%%%%%%%%%%%%%%
\section{The model}
We will study the dynamics of a model for two single-mode Bose--Einstein
condensates  
which are coupled via Josephson tunneling. The model is not only
applicable to tunneling in atomic Bose--Einstein condensates, but also 
mesoscopic solid state Josephson junctions \cite{leg,int8,mss} and non-linear
optics \cite{saf}. The Hamiltonian is
given by 

\begin{equation}
H=\frac{k}{8}(N_1-N_2)^2 - \frac{\Delta \mu}{2}(N_1-N_2)
- \frac{{\cal E}_\j}{2}(a^{\dag}_1 a_2 +a_2^{\dag} a_1)
\label{ham}
\end{equation}
where we follow the notational conventions of \cite{leg} for the
couplings. Above,  $a^{\dag}_i, a_i$ ($i=1,2$) are the creation and
annihilation operators for bosons occupying one of two modes,
labelled 1 and 2, and $N_1$ and $N_2$ are the respective number
operators. The operator $N=N_1+N_2$ is the total boson number operator
and it is conserved. The coupling $k$ provides the strength of the
scattering interaction between bosons, $\Delta \mu$ is the
external potential and ${\cal E}_\j$ is the coupling for the
tunneling. We remark that a change of sign in the tunneling
coupling ${{\cal E}_\j}\rightarrow {-{\cal E}_\j}$ is equivalent
to the unitary transformation 
$ a_1\rightarrow a_1,\,a_2\rightarrow
-a_2$.     
In that which follows we will mostly discuss the case $\Delta\mu=0$.

Let us now, for future use, distinguish 
different coupling regimes of the model characterised by different
ratios of
$k/{\cal E}_\j$. We first mention that there 
exists a threshold coupling $k/{\cal E}_\j=4/N$ 
which signifies the transition between delocalisation
and self-trapping
\cite{milb} (see also \cite{mjcz}):   

\begin{itemize}
\item Delocalisation $\longrightarrow$  $k/{\cal E}_\j < 4/N$

\item Self-trapping $\longrightarrow$  $4/N < k/{\cal E}_\j$

\end{itemize}
Following \cite{leg}, it is also useful to consider the following  
three 
regimes:  

\begin{itemize}
\item  Rabi  regime $\longrightarrow$  $k/{\cal E}_\j << 1/N$

\item Josephson  regime $\longrightarrow$  $1/N << k/{\cal E}_\j << N$

\item Fock  regime $\longrightarrow$  $k/{\cal E}_\j >> N.$

\end{itemize}
For these three regimes it is known that an analogy can be drawn between 
the dynamics of (\ref{ham}) and the dynamics of certain pendulums \cite{leg}. 
Comparing the above classifications it is seen that the threshold  
coupling lies in the crossover between the Rabi and Josephson regimes,
thus offering a candidate to precisely define the boundary between
these regimes. 

Below the quantum dynamics of (\ref{ham}) will
be investigated using the method of direct numerical
diagonalisation, which allows
for a study across all couplings.
It is well known that the time evolution of any physical quantity
is determined through the temporal
operator $U$, given by

\begin{equation}
U = \sum_{n=0}^N e^{-i\lambda_n t}|\psi_n \rangle \langle\psi_n|
\label{ut}
\end{equation}
where $\{\lambda_n\}$ and  $\{|\psi_n \rangle\}$
are the eigenvalues and eigenvectors of the Hamiltonian (\ref{ham}).
Therefore, the temporal evolution of any state can be calculated by

\begin{equation}
|\psi (t)\rangle = \sum_{n=0}^N a_n  e^{-i\lambda_n t}|\psi_n\rangle,
\label{pt}
\end{equation}
where  $a_n = \langle \psi_n|\phi \rangle$ and $|\phi \rangle$
represents the initial state. Using eqs. (\ref{ut}) and (\ref{pt})
we can compute the expectation value and the variance (which is a measure of 
the quantum fluctuations) 
of any physical quantity represented by a self-adjoint operator,
as well as the entanglement. In particular, the temporal
dependence of the expectation value of any operator $A$ can be
computed using the expression

$$
\langle A \rangle = \langle \psi(t)|A|\psi(t)\rangle,
$$
while the variance is given by
$$
\Delta(A) = \langle A^2 \rangle  - \langle A \rangle^2.
$$

Following \cite{bbps}, for any pure state density
matrix $\rho$ of a bipartite system the entanglement is defined by
$$E(\rho)=-{\rm tr}(\rho_1 \log_2 \rho_1)=-{\rm tr}(\rho_2 \log_2 \rho_2)$$
where ${\rho}_{1}$ is the reduced density matrix obtained by 
taking the partial trace of $\rho$ over the subsystem
$2$. The definition for $\rho_2$ is analogous.  
Here we follow the approach argued in \cite{hmm}. Due to 
indistinguishability, it is not useful to consider entanglement between
individual bosons. Rather, we take the two modes of the Hamiltonian (\ref{ham})
to define 
two subsystems, which can be distinguished through the operators $N_1$
and $N_2$  representing measurement of each subsystem population.
In this case, $E(\rho)$ is equivalent to \cite{hmm}
$$
E(\rho ) = -\sum_{n=0}^{N} |c_n(t)|^2\log_2(|c_n(t)|^2)
$$
where the $c_n(t)$ are the co-efficients of a general state of the
system

$$
|\psi(t) \rangle = \sum_{n=0}^N c_n(t) |N-n,n \rangle.
$$
The states $|m,n \rangle $,  which collectively provide a basis
for the Fock space,
are given by

$$
|m, n \rangle =\frac{(a_1^{\dagger})^{m}(a_2^{\dagger})^{n}}
{\sqrt{m!}\sqrt{n!}}|0 \rangle ,
$$
and $|0 \rangle $ is the Fock space vacuum. For a system of $N$
total particles the maximal entanglement is $E_{\rm
max}=\log_2(N+1)$. 

\vspace{1.00cm} 
\begin{figure}[ht]
\begin{center}
\epsfig{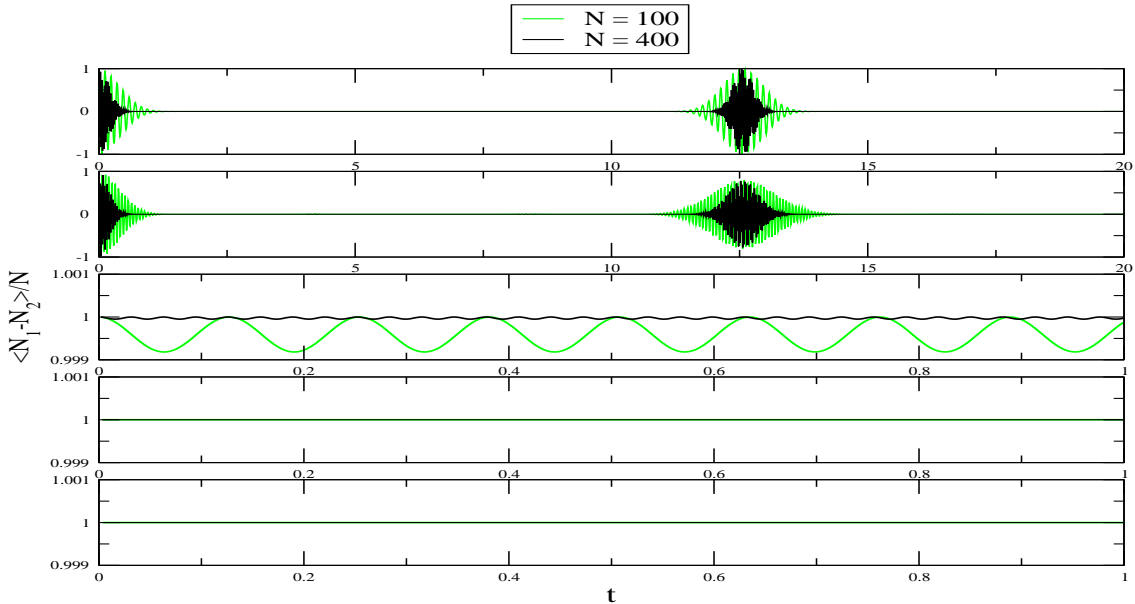}
\caption{Time evolution of the expectation value for the relative
number of particles
for different ratios of the coupling
$k/{\cal E}_\j$ from the top (Rabi regime) to the bottom (Fock
regime):
$k/{\cal E}_\j= 1/N^2,1/N,1,N,N^2$ for $N=100, 400$ and the initial
state is $|N,0\rangle $.}
\label{fig.2}
\end{center}
\end{figure}

\section{Expectation value of the relative particle number}

First we investigate the quantum dynamics of the relative particle
number operator $N_1-N_2$ (or the {\emph {imbalance}}), in the
three different regimes, Rabi, Josephson and Fock, as well as in
the intermediate ones. For this purpose we use the expressions of
the previous section together with the eigenvalues and
eigenvectors obtained directly from numerical diagonalisation of
the Hamiltonian (\ref{ham}).

{}From fig. \ref{fig.2}, it is apparent that the qualitative behaviour in
each region, using the same initial state, does not depend on the
number of particles. (In this figure, as in the others following,
the chosen time interval is one which most clearly shows the
relevant dynamical behaviour.) We found that in the interval
$k/{\cal E}_\j\in [1/N^2,1/N]$ (close to the Rabi regime) the
collapse and revival time takes the constant value
$t_{cr}=4\pi$\footnote{The ratio $k/{\cal E}_\j=1/N^2$ means that
we are using $k=1$ and ${\cal E}_\j=N^2$ and similarly for the
other cases. This convention, which will be used throughout the
paper unless noted otherwise, fixes the time scale.}. 
Also, in the
region where $k/{\cal E}_\j-1$ is small and negative, the expectation value 
still displays collapse and revival behaviour with  
harmonic oscillations occurring at
the value $k/{\cal E}_\j=1$. 
At this coupling the period of oscillations
is not independent of $N$. Still with reference to fig.
\ref{fig.2}, we observed in the region where $k/{\cal E}_\j-1$ is
small and positive that the expectation value has small amplitude periodic
oscillation close to the initial expectation value of $1$. The
amplitude of the oscillations disappear as the ratio  $k/{\cal E}_\j$ becomes
larger (near the Fock regime) at which point the bosons remain
localised to the subsystem of the initial configuration. However
this transition  from small amplitude harmonic oscillations to
complete localisation appears to be smooth. Hereafter we will
focus most of our attention to the Rabi and Josephson regimes,
including the crossover, as this is where the most complex
behaviour is to be found.

\vspace{1.0cm}
\begin{figure}[ht]
\begin{center}
\epsfig{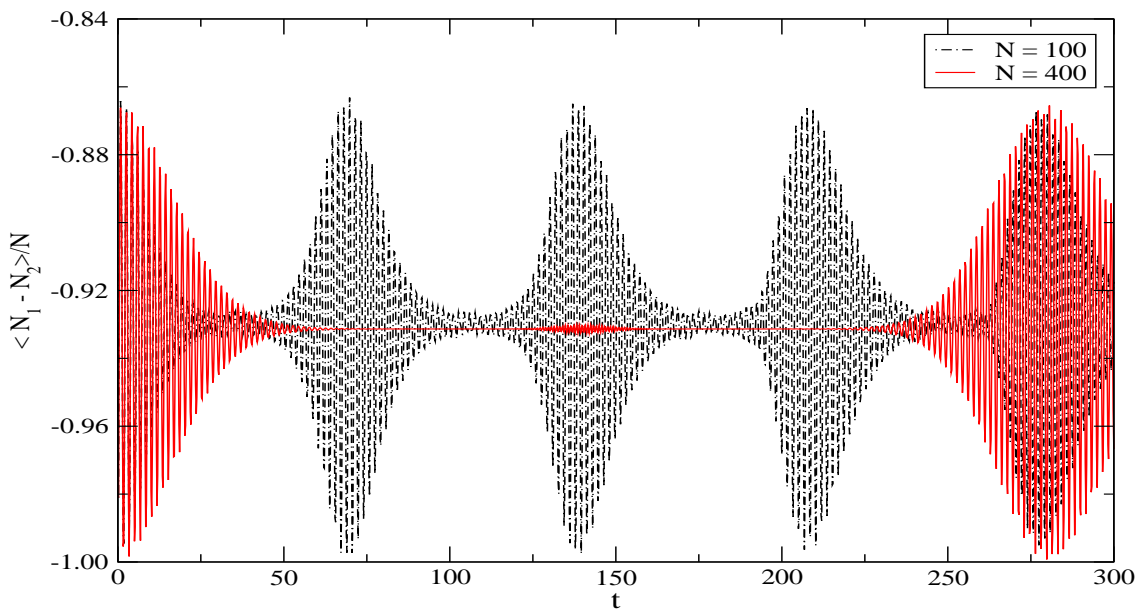}
\caption{Time evolution for the expectation value
of the relative number of particles for the initial state
$|0,N\rangle$. The dot-dashed line is for $N=100$ and the solid
line is for $N=400$ with ${\cal E}_\j =1$ and $k=8/N$.}
\label{fig.1}
\end{center}
\end{figure}

In fig. \ref{fig.1}, we plot the expectation value for the
relative number of particles for two different cases: $N=100$ and
$N=400$, with the initial state $|0,N\rangle$. Using the relation
$k/{\cal E}_\j =8/N$ we reproduce figure 5 of \cite{milb}, taking
into account the different notational conventions and using a
different time interval to show the collapse and revival for both
cases. For this case we have followed the conventions used in
\cite{milb} for the time scale by setting ${\cal E}_\j=1 $ and
$k=8/N$. It is apparent that under this convention the collapse
and revival time is not independent of $N$. However, introducing a
rescaled collapse and revival time $t_{cr}^*=8t_{cr}/N$, which
corresponds to our adopted time scale convention, shows that
$t_{cr}^*$ is independent of $N$. A distinguishing feature of fig.
\ref{fig.1} is that the mean value (i.e. time-averaged value) 
of the imbalance is not zero,
in contrast to the collapse and revival sequences shown in fig.
\ref{fig.2}. 
This is because the respective couplings lie in regimes 
on different sides of the
threshold value $k/{\cal E}_\j =4/N$, 
and these cases respectively illustrate    
the behaviour associated with self-trapping and delocalisation. 

\begin{figure}[ht]
\begin{center}
\epsfig{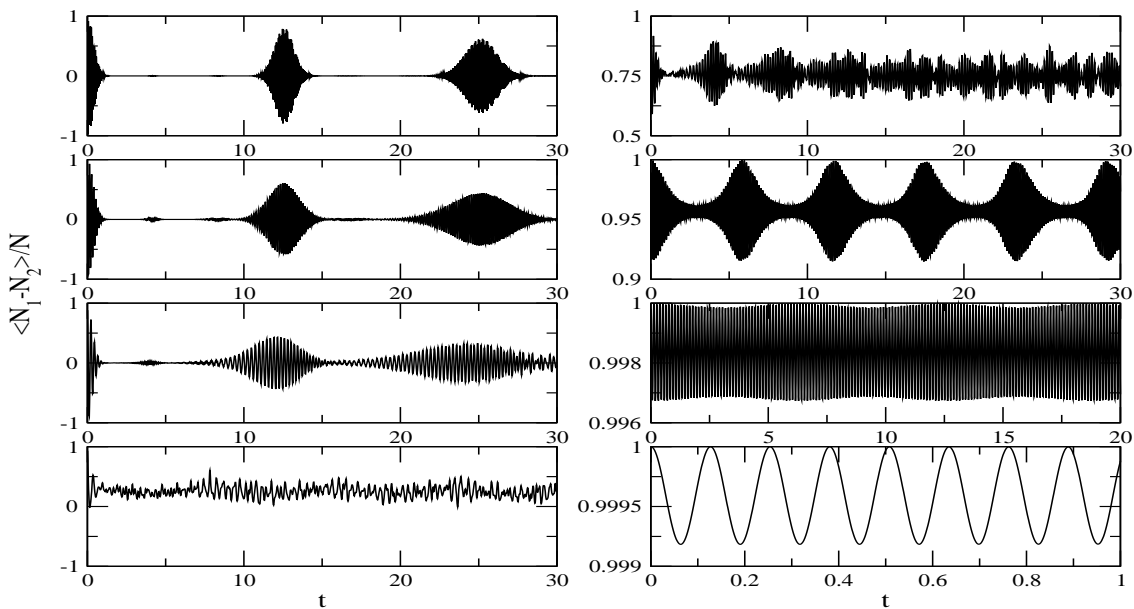}
\caption{Time evolution of the expectation value between
$k/{\cal E}_\j=1/N$
and  $k/{\cal E}_\j= 1$. On the left, from top to bottom
$k/{\cal E}_\j= 1/N,2/N,3/N,4/N$ and on
the right, from top to bottom  $k/{\cal E}_\j=5/N,10/N,50/N,1$, where
 $N=100$ and  the initial state is $|N,0\rangle $.}
\label{fig.3}
\end{center}
\end{figure}

In fig. \ref{fig.3} we can see in detail for the case $N=100$ 
the evolution of the dynamics from a collapse and revival sequence with
$t_{cr}=4\pi$ for $k/{\cal E}_\j<4/N$, through  the self-trapping
transition at $k/{\cal E}_\j=4/N$, and toward small amplitude
harmonic oscillations in the imbalance of the localised state when
$k/{\cal E}_\j=1$. 
Where there is clear collapse and revival
in the self-trapping phase we find that $t_{cr}$ increases with
increasing $k/{\cal E_\j}$. Further increases in $k/{\cal E_\j}$
lead to a decaying of the collapse and revival sequence toward
harmonic oscillations which occur at $k/{\cal E}_\j=1$.
\vspace{1.0cm}
\vspace{1.0cm}
\begin{figure}[ht]
\begin{center}
\epsfig{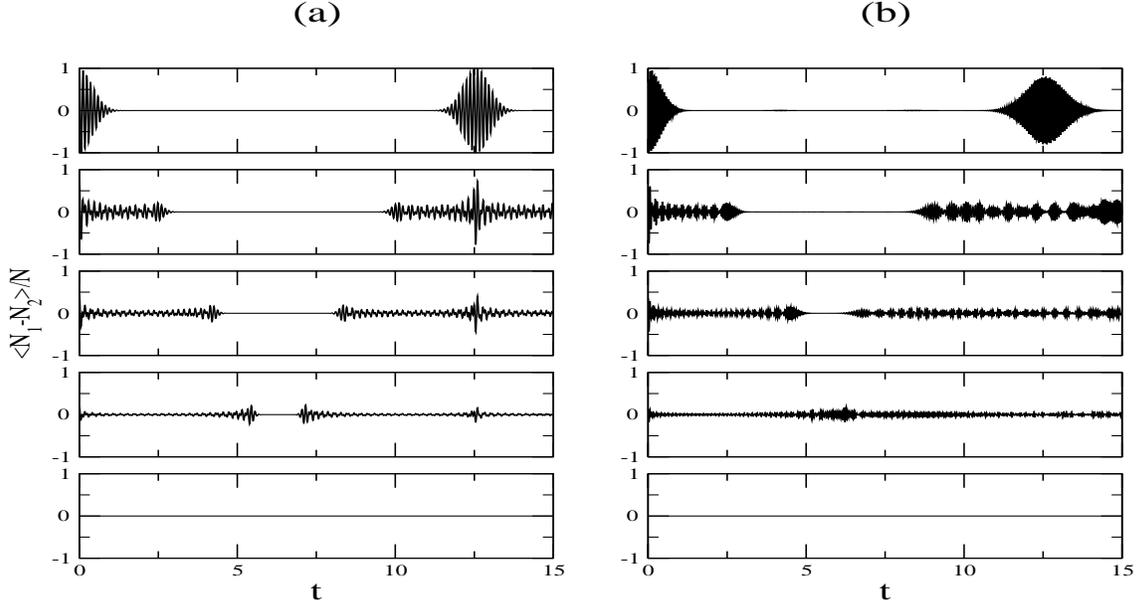}
\caption{ Relative number expectation values for
(a) the Rabi  regime ($k/{\cal E}_\j=1/N^2$);
(b) between the Rabi and Josephson
regimes ($k/{\cal E}_\j=1/N$). The initial states, from top to bottom
are
$|100,0\rangle ,
|90,10\rangle ,|74,26\rangle ,|60,40\rangle $ and $|50,50\rangle $.}
\label{figa}
\end{center}
\end{figure}

Next we turn our attention to study the evolution of the
expectation value $\langle N_1-N_2 \rangle /N$ using a range of
different initial conditions. In the first of these studies fig.
\ref{figa}, we illustrate the types of different behaviours that
occur by using different initial states. Here, it is clear that
the collapse and revival time depends on the choice of initial
state. As expected the mean amplitude of the oscillations is
related to the imbalance of the initial state: the mean amplitude is
maximal when the initial state has all particles in the same
subsystem ($|N,0\rangle$ and $|0,N\rangle$), and it is minimal
when the initial state has the same number of particles in each
subsystem ($|N/2,N/2\rangle$). In both regimes shown in fig.
\ref{figa} the expectation value of the relative number of
particles is symmetric about zero, with oscillations in the
relative number disappearing when the imbalance of the initial
state is zero. 
%\vspace{1.0cm}
\begin{figure}[ht]
\begin{center}
\epsfig{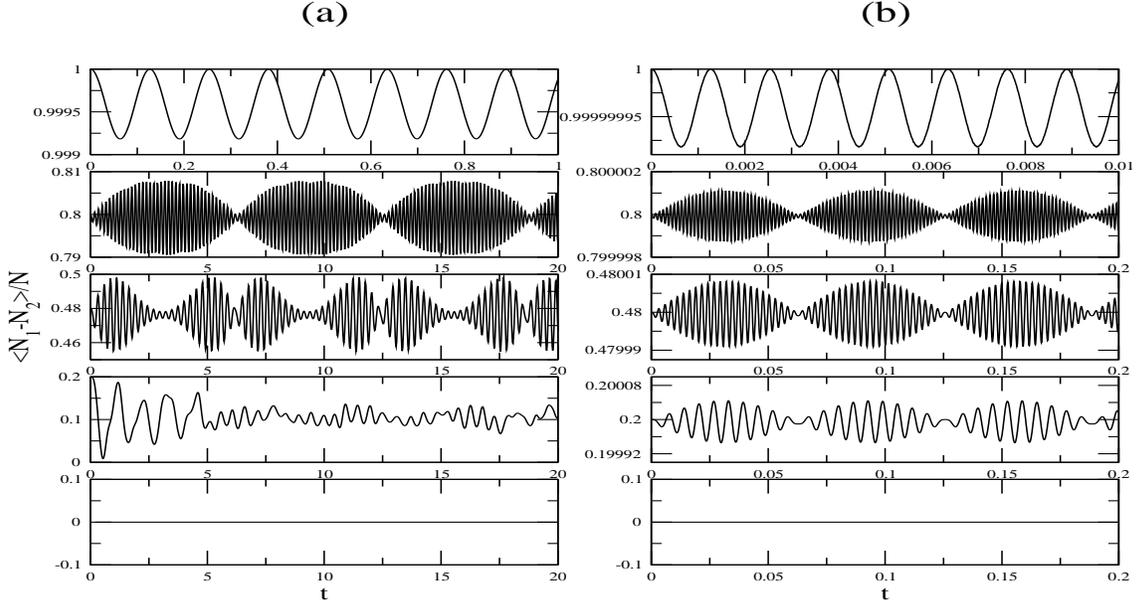}
\caption{Relative number expectation values for (a) the Josephson
regime ($k/{\cal E}_\j= 1$); (b) between the Josephson and Fock
regimes ($k/{\cal E}_\j=N$). The initial states are, from top to
bottom: $ |100,0\rangle ,|90,10\rangle ,|74,26\rangle
,|60,40\rangle $ and $|50,50\rangle$.} \label{figb}
\end{center}
\end{figure}

As we vary the coupling ratio 
past the self-trapping threshold value 
we find an entirely different scenario. In fig. \ref{figb} the
expectation value for the relative number of particles shows a
variety of different behaviours. The dynamics is harmonic when the
initial state corresponds to all particles in the same subsystem,
and the oscillation amplitude is small. As the imbalance in the
initial state decreases, we first see harmonic modulation of the amplitude of
the oscillations,
in stark contrast to the cases shown in fig. \ref{figa}. Note here
that the period of the modulation does not depend on the initial
state. 
It is important to mention that in the interval $k/{\cal
E}_\j\in [1/N,1]$, which is not shown, we find that the crossover from 
collapse and revival sequences of oscillations with zero mean to 
oscillations with mean approximately equal to the initial imbalance
occurs at $k/{\cal E}_\j=4/N$, as expected. 

When using other initial states, we can classify them into two
categories: symmetric,  such as a cat-state
\begin{equation}
|\phi_{\rm CAT}\rangle=\frac{1}{\sqrt{2}}(|N,0\rangle +|0,N\rangle
) \label{cat} \end{equation} and a maximally entangled state
\begin{equation}
|\phi_{\rm ME}\rangle=\frac{1}{\sqrt{N+1}}\sum_{n=0}^{N}|N-n,n
\rangle , \label{me} \end{equation} or non-symmetric, such as
$|N,0\rangle ,|N-10,10\rangle ,$ etc. The expectation value for
the number of particles of any symmetric initial state is zero,
due to symmetry of the Hamiltonian under interchange of the two
subsystems. In such a case some insight into the behaviour of the
system can be gleaned by investigating the variance of the
imbalance, which will be explored in the next section.

%%%%%%%%%%%%%%%%%%%%%%%%%%%%%%%%

When a non-zero external potential $\Delta\mu$ is added to the
system, some new features appear in the various regimes. In the
extreme Rabi regime, the imbalance population oscillates around
zero, similar to the $\Delta\mu =0$ case, but here the dependence on the
initial
state is different and in particular the oscillation amplitude is
not zero using the initial state $|N/2,N/2\rangle$. 
Another important point is that
the initial states $|n_1,n_2\rangle$ and $|n_2,n_1 \rangle$
produce the same expectation value behaviour. In the
crossover between Rabi and Josephson regimes, a different behaviour
is found: the oscillation amplitude of the number of particles
also depends on the initial state, but here the initial states
$|n_1,n_2\rangle$ and $|n_2,n_1\rangle$ do not produce the same
expectation value behaviour. The symmetry breaking of the Hamiltonian
due to the
term $\Delta\mu$ is prominent in this case.

In the Josephson and Fock regimes the introduction of the external
potential only produces a significant difference in comparison to
the $\Delta\mu =0$ case when the initial imbalance is small. If we
consider an initial state where the imbalance population is large,
then the two initial states $|m,n\rangle$ and $|n,m\rangle$ produce
essentially the same results, while when $|m-n|$ is close to zero
the same initial states produce different results. This aspect can
be understood when we write down the first two terms of the
Hamiltonian (\ref{ham}) in the following way:

$$\left(\sqrt{\frac{k}{8}}(N_1-N_2)-\sqrt{\frac{1}{2k}}\Delta\mu\right)^2
-\frac{1}{2k}\Delta\mu^2.$$
When the fluctuations in the imbalance are small, we see from the
terms in the brackets that the external potential will be dominant
when

$$\Delta\mu > \frac{k}{2}\langle N_1-N_2\rangle .$$
This means that the largest effect of the external potential will
generally occur when the initial state is symmetric.

\section{Imbalance fluctuation and entanglement}

Here we will  study the extent to which the evolution of the imbalance
variance and the entanglement are correlated, and how the choice of 
initial state, as well as the coupling regime, determine the main features 
of the evolution. 
On one hand, any  state of the system (with fixed $N$) 
which is unentangled must
have zero fluctuation in the imbalance. The converse however is not
true, as the cat-state (\ref{cat}) has maximal imbalance variance but
relatively small entanglement, and the maximally entangled state (\ref{me})
has smaller imbalance variance than (\ref{cat}).  
%\vspace{1.00cm}
\begin{figure}[ht]
\begin{center}
\epsfig{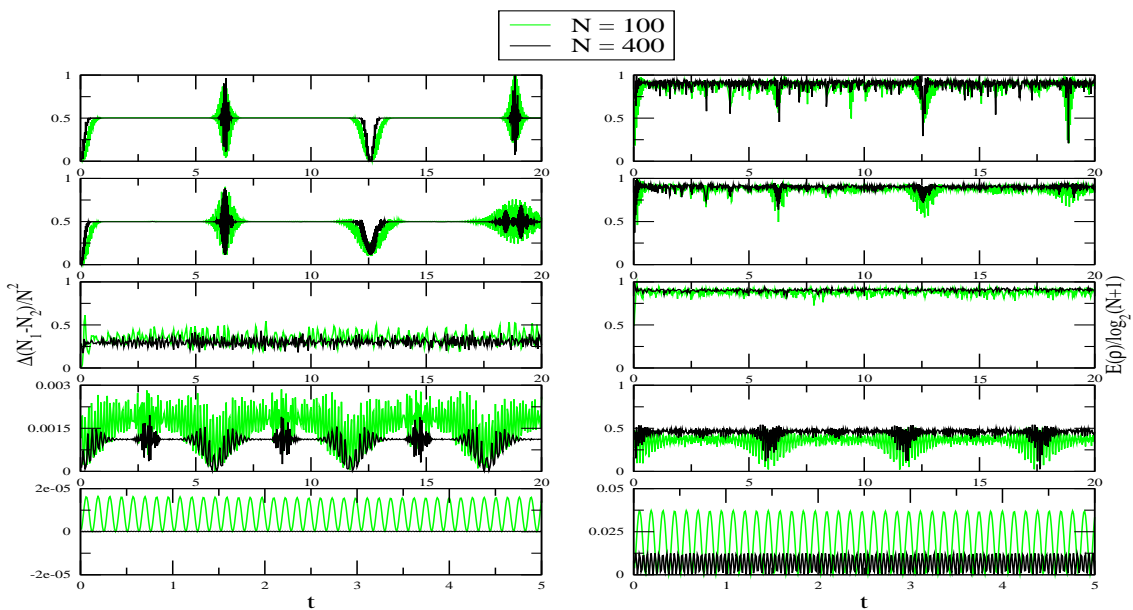}
\caption{Time evolution of the relative particle number
fluctuations (on the left) and entanglement (on the right) for the
initial state $|N,0\rangle $ where $N=100,\,400$. From
 top to bottom the ratio of couplings used are 
 $k/{\cal E}_\j=1/N^2,1/N,4/N,10/N,1$.} \label{fig.4}
\end{center}
\end{figure}

For the results shown in fig. \ref{fig.4}, we have chosen $|N,0
\rangle $ as the initial state, which is unentangled. In the coupling
regime $k/{\cal E}_\j\in [1/N^2,1/N]$ 
where the tunneling is strong, we find that the behaviour
of the variance of the relative particle number is similar to
the collapse and revival character of the expectation value (cf.
fig. \ref{fig.2}). We
can observe that the entanglement quickly approaches the maximum
value and  for all later times the system is mostly entangled. 
At the threshold coupling $k/{\cal E}_\j=4/N$, there is no collapse and revival 
behaviour in the imbalance variance and we see that the mean value starts to 
decrease.  
Beyond the threshold coupling, the 
mean value of both the imbalance variance and the entanglement decrease.    
When the ratio $k/{\cal E}_\j=1$ is
reached, we find  small oscillations close to zero for both the
imbalance variance and the entanglement. From here to the Fock regime (not
shown), the imbalance variance practically goes to zero and the
entanglement essentially disappears. 
\vspace{0.5cm}
\begin{figure}[ht]
\begin{center}
\epsfig{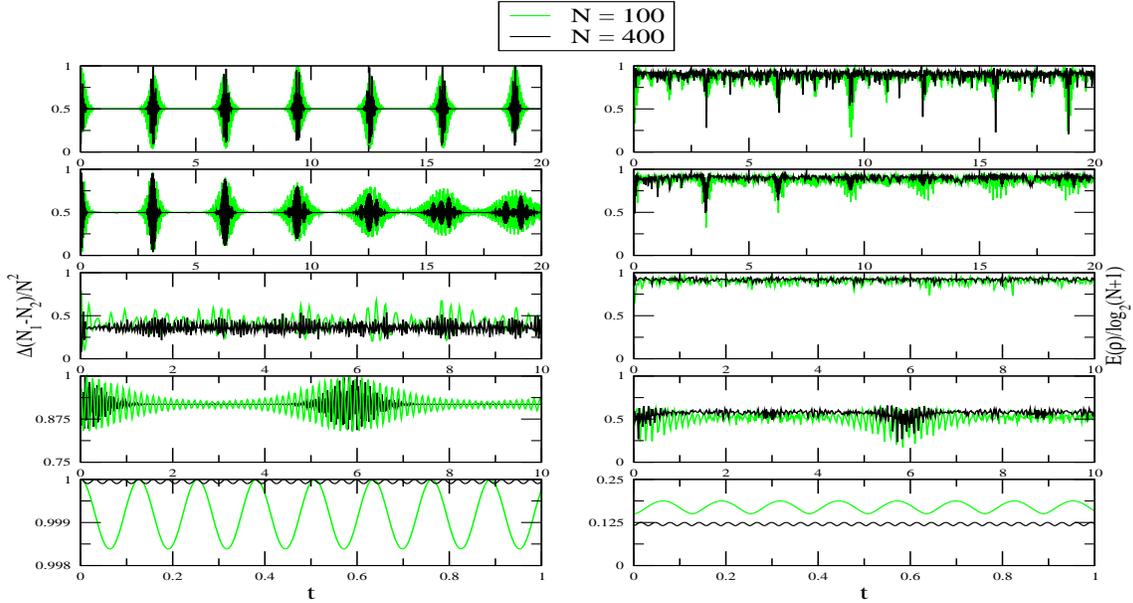}
\caption{Time evolution of the relative particle
number fluctuations (on the left) and entanglement (on the right)
for the initial state $|\phi_{\rm CAT}\rangle $ where
$N=100,\,400$. From top to bottom the ratio of couplings used are
$k/{\cal E}_\j=1/N^2,1/N,4/N,10/N,1$.} \label{fig.5}
\end{center}
\end{figure}

For fig. \ref{fig.5}, the initial state is the cat-state
(\ref{cat}) which has entanglement $E(\rho)=1$.  
In contrast to fig. \ref{fig.4}, the mean value of the 
imbalance variance increases beyond the threshold coupling.  
As the ratio $k/{\cal
E}_\j$ approaches unity, small amplitude periodic oscillations are
observed very near to the maximal value of $N^2$. For the
entanglement the picture is similar to fig. \ref{fig.4}. 
For $k/{\cal
E}_\j=1$ the entanglement is found to display periodic
oscillations about a mean value close to the value for the 
entanglement of the initial state, which is non-zero in this instance. 
%\vspace{1.0cm}
\begin{figure}[ht]
\begin{center}
\epsfig{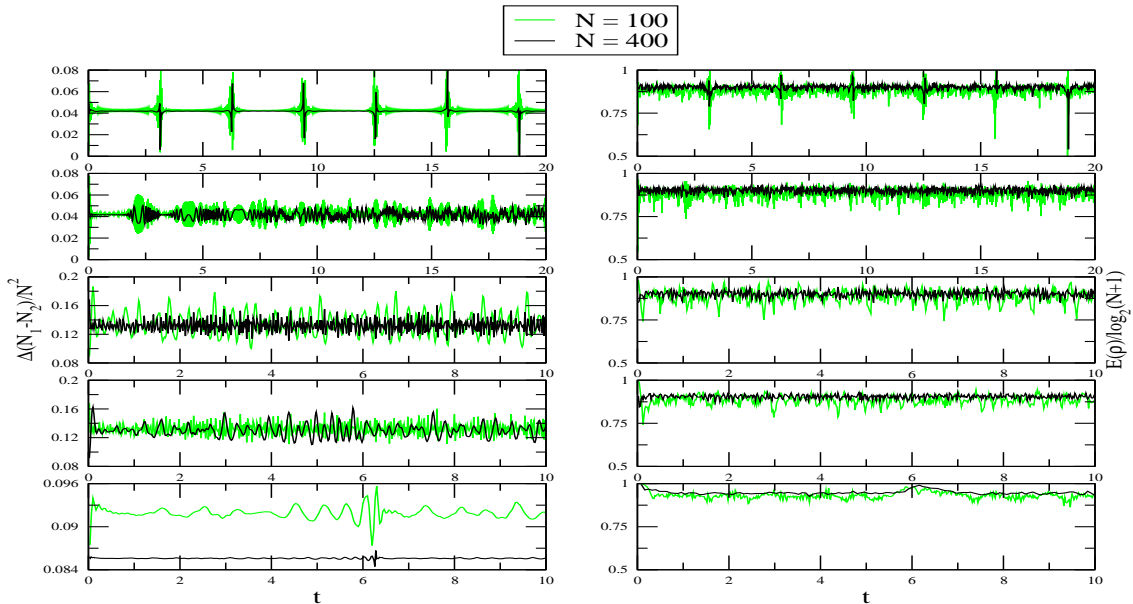}
\caption{Time evolution of the relative particle
number fluctuations (on the left) and entanglement (on the right)
for the initial state $|\phi_{\rm ME}\rangle $ where
$N=100,\,400$. From
 top to bottom the ratio of couplings used are 
 $k/{\cal E}_\j=1/N^2,1/N,4/N,10/N,1$.} \label{fig.6}
\end{center}
\end{figure}

Finally, fig. \ref{fig.6} shows the results obtained when the
initial state is the maximally entangled state (\ref{me}). Here, a
much different behaviour is found in comparison to the previous
examples. In the interval $k/{\cal E}_\j\in [1/N^2,1/N]$ the mean
values for the variance of the imbalance are \emph{smaller} 
than in the previous two examples. 
At the threshold coupling the mean is substantially larger than the mean value 
below threshold coupling, and the mean value again decreases as 
$k/{\cal E}_\j$ approaches unity. The most striking feature at 
$k/{\cal E}_\j=1$ is 
that there is no discernable periodicity in the imbalance variance, 
whereas in the analogous cases
shown in figs. \ref{fig.4}, \ref{fig.5} there is clear periodicity.
For the entanglement, it appears that the evolution is not substantially 
dependent on the coupling regime when the initial state is maximally entangled. All cases shown have much the same mean entanglement over time. Like the 
imbalance variance, there is also no periodicity in the entanglement evolution 
for the coupling 
$k/{\cal E}_\j=1$.

%%%%%%%%%%%
\section{Conclusion}
To summarise, we have investigated the quantum dynamics of a model for
two Josephson-coupled Bose--Einstein condensates across a wide range of 
coupling regimes and various initial states,  
using the method of direct numerical diagonalisation.
Our analysis shows that despite the apparent simplicity of the
Hamiltonian, 
diverse features including 
collapse and revival of oscillations, amplitude modulated oscillations
and non-harmonic behaviour are displayed.  
The most striking changes in the dynamics occur as the system crosses
the threshold coupling $k/{\cal E}_\j=4/N$.  

\vspace{2.0cm}
\centerline{{\bf Acknowledgements}}
~\\
APT thanks CNPq-Conselho Nacional de Desenvolvimento
Cient\'{\i}fico e Tecnol\'ogico (a funding agency of the Brazilian Government)
for financial support and the Centre for Mathematical Physics at
The University of Queensland for kind of hospitality. JL
gratefully acknowledges funding from the Australian Research Council 
and The University of Queensland through a
Foundation Research Excellence Award. AF thanks CNPq-Brazil.

%%%%%%%% %


\begin{thebibliography}{99}
\bibitem{bose} S. N. Bose, Z. Phys.  {\bf 26} (1924) 178
\bibitem{eins} A. Einstein, Phys. Math. K1  {\bf 22} (1924) 261
\bibitem{cw} E. A. Cornell and C. E. Wieman, Rev. Mod. Phys. {\bf 74}
(2002) 875
\bibitem{ak} J. R. Anglin and W. Ketterle, Nature {\bf 416} (2002) 211
\bibitem{int3} M. R. Andrews, C. G. Townsend, H. -J. Miesner,
D. S. Durfee, D. M. Kurn and W. Ketterle, Science  {\bf 275} (1997) 637
\bibitem{dctw} E. A. Donley, N. R. Claussen, S. T. Thompson and C. E.
Wieman, Nature {\bf 417} (2003) 529
\bibitem{y} L. You, Phys. Rev. Lett. {\bf 90} (2003) 030402
\bibitem{nc} M. A. Nielsen and I. L. Chuang, {\it Quantum Computation
and Quantum Information} (Cambridge University Press, Cambridge, 2000)
\bibitem{intaa} J. Willians, R. Walser, J. Cooper, E. A. Cornell, M.
Holland,
Phys. Rev. A {\bf 61} (2000) 033612
\bibitem{intbb} M. R. Mathews, B. P. Anderson, P. C. Haljan, D. S. Hall,
M. J. Holland, J. E. Willians, C. E. Wieman and E. A. Cornell, Phys.
Rev. Lett.
{\bf 83} (1999)  3358.
\bibitem{gmehb} M. Greiner, O. Mandel, T. Esslinger, T.W. H\"ansch and
I. Bloch, Nature {\bf 415} (2002) 39
\bibitem{leg} A. J. Leggett, Rev. Mod. Phys. {\bf 73} (2001) 307
\bibitem{milb} G. J. Milburn, J. Corney, E. M. Wright and D. F. Walls,
Phys. Rev. A  {\bf 55} (1997) 4318
\bibitem{hmm} A. P. Hines, R. H. McKenzie and G. J. Milburn, Phys. Rev.
A {\bf 67} (2003) 013609
\bibitem{int8} J. R. Anglin, P. Drummond, A. Smerzi, Phys. Rev. A
{\bf 64} (2001) 063605
\bibitem{lz} J. Links and H.-Q. Zhou, Lett. Math. Phys. {\bf 60} (2002)
275
\bibitem{zlmg} H.-Q. Zhou, J. Links, R. H. McKenzie and X. -W. Guan,
J. Phys. A: Math. Gen.  {\bf 36} (2003) L113
\bibitem{mjcz} A. Micheli, D. Jaksch, J. I. Cirac and P. Zoller, Phys.
Rev. A {\bf 67} (2003) 013607
\bibitem{on} T. J. Osbourne and M. A. Nielsen, Phys. Rev. A {\bf 66}
(2002) 032110
\bibitem{oaff} A. Osterloh, L. Amico, G. Falci and R. Fazio, Nature {\bf
416} (2002) 608
\bibitem{md} M. A. Mart\'{\i}n-Delgado, preprint quant-ph/0207026
\bibitem{saf} L. Sanz, R. M. Angela and K. Furaya, J. Phys. A: Math.
Gen. {\bf 36} (2003) 9737
\bibitem{mss} Y. Makhlin, G. Sch\"on and A. Shnirman, Rev. Mod. Phys.
{\bf 73} (2001) 357 
\bibitem{bbps} C. H. Bennett, H. J. Bernstein, S. Popescu and B.
Schumacher, Phys. Rev. A {\bf 53} (1996) 2046
\end{thebibliography}
 \end{document}